\documentclass[aps,prd,10pt,tightenlines,amsmath,unsortedaddress,twocolumn,amssymb,10pt]{revtex4}

\usepackage{float}
\usepackage{graphicx}
\usepackage{dcolumn}
\usepackage{bm}
\usepackage{bbold}
\usepackage{enumerate}
\usepackage[utf8]{inputenc}
\usepackage[english]{babel}
\usepackage{amsthm}

\usepackage{xcolor}

\usepackage[breaklinks,pdftex]{hyperref}

\begin{document}

\title{Impact of local foreground radiations on Very High Energy observations of extragalactic sources}

\author{Alberto Franceschini}
\email{alberto.franceschini@unipd.it}
\affiliation{University of Padova, Physics and Astronomy Department, Vicolo Osservatorio 3, I -- 35122 Padova, Italy}
\affiliation{INAF, Osservatorio Astronomico di Brera, Via E. Bianchi 46, I -- 23807 Merate, Italy}

\author{Giorgio Galanti}
\email{gam.galanti@gmail.com}
\affiliation{INAF, Istituto di Astrofisica Spaziale e Fisica Cosmica di Milano, Via A. Corti 12, I -- 20133 Milano, Italy}

\date{\today}

\begin{abstract}
In the framework of exploiting Very High Energy (VHE) gamma-ray observations of extragalactic sources to infer constraints on the intensity of cosmological background radiations, as well as on deviations from the standard model of particles and physical interactions (like violations of Lorentz invariance and axion-like particles -- LIV and ALPs), we discuss here possible contaminant effects due to the presence of radiations fields from local sources. We specifically model and analyze the foreground radiations produced inside the host galaxy of the VHE source and the cosmic environment itself hosting the source, as well as radiations produced in the Milky Way, along the VHE source's observational line-of-sights. Our analysis shows that such contaminant foregrounds may indeed impact on observations of only the very local Active Galactic Nuclei (e.g. Centaurus A), but not significantly those in the Virgo cluster (M87) and beyond. 
 
\end{abstract}


\maketitle

%

\section{Introduction}

\label{S1}

Detecting Very High Energy (VHE) photons from extreme astrophysical sources in the extragalactic space offers important tests for physics and astrophysics.
These rely on the quantum-mechanical process of physical interaction between VHE photons emitted by cosmic sources and the very numerous lower-energy diffuse background photons. Such interactions destroy the two incident photons by generating an electron-positron pair.
Among these tests, one important is the possibility to verify deviations from the standard model of particles and physical interactions, that are expected to occur at VHE, like violations of the Lorentz Invariance (LIV, see \cite{addazi2022}), the mixing of VHE photons and axion-like particles (ALPs, see \cite{grRev}), and others.

At the same time, interesting constraints in the field of astrophysics and cosmology can be obtained from such VHE observations, thus creating a bridge between fundamental physics and astrophysics. The latter concern, for example, the spectral intensities of the diffuse radiation backgrounds of cosmological origin from the optical-UV to the millimeter, the so-called Extragalactic Background Light (EBL).
Various attempts on this regard have been reported, the vast majority concerning so far the EBL section from $\lambda=0.1$ to 10 $\mu {\rm m}$ \cite{2006Natur.440.1018A}. This is indeed the range where the highest energy interacting photons fall in the maximal sensitivity interval, around 1 TeV, for the current Cherenkov telescope arrays, given the approximate relation for the maximum of the photon-photon interaction probability:
\begin{equation}\label{eq1}
\lambda \ [\mu {\rm m}]\simeq 1.38 \ E \ [\rm TeV].
\end{equation}

On the contrary, the infrared section of the EBL from about $\lambda=10\ \mu {\rm m}$ to the sub-millimeter, is essentially currently almost unconstrained, because it would require to extend the sensitivity for extragalactic sources above the present 10 TeV limit, with one exception based on difficult VHE observations of a GRB event \cite{2023SciA....9J2778C}.
Observing above this limit will become feasible with the forthcoming new generation instrumentation 
(ASTRI Mini Array \cite{2022JHEAp..35...52S}, CTAO \cite{2011ExA....32..193A},  LHAASO \cite{2019ChA&A..43..457C}).

In more detail, the test's procedure consists in comparing the VHE spectral observations, $F_{\rm obs}$, of distant sources with emission model expectations, $F_{\rm mod}$, to infer the gamma-gamma optical depth $\tau_{\gamma \gamma}$:
\begin{equation}\label{eq1a}
 F_{\rm obs}=F_{\rm mod} \exp(-\tau_{\gamma \gamma}),
\end{equation}
where in turn, in the simplest approximate terms,
\begin{equation}\label{eq2}
\tau_{\gamma \gamma} \simeq \sigma_{\gamma \gamma} n_\gamma \ell .
\end{equation}
Here $\sigma_{\gamma \gamma}\simeq 0.4\sigma_T$ is the photon-photon cross section with the maximum given by Eq. \ref{eq1}, $n_\gamma$ is the background low-energy photon field density in ${\rm cm}^{-3}$, $\ell$ is the length of the photon interaction region, or the source distance when the EBL absorption is considered.

From Eq. \ref{eq2} it is clear that the gamma-gamma interaction effects become appreciable only in relation to the long distances of extragalactic sources, particularly when attempting to constrain the EBL intensity or the effects of the photon-ALP mixing.
At the same time, if the distance $\ell$ to the VHE source is too large, then all the highest energy photons are immediately lost from the spectrum, so that a delicate optimization of the target objects, typically to be found among those at low-redshifts, is required.

It is also clear that, when looking at very low-$z$ sources, problems can arise from the fact that the analysis of the photon-photon interaction may become confused by the variety of radiation fields that the VHE photons happen to cross, in particular the local foregrounds in the source itself, those in the structure hosting the source (like the host galaxy cluster or group), and in the local reference system (the Galaxy).
All these are discussed in the present paper.

The latter is structured as follows. Sec. \ref{S2} reports a summary of the procedure that we adopted to calculate the optical depth for the gamma-gamma interactions.
In Sec. \ref{S3} we review the case of using photon-photon opacity measurements at high VHE energies to constrain the Extragalactic Background at long IR wavelengths, which is currently largely unconstrained.
In Sec. \ref{S4} we calculate the contributions to the opacities by local radiation foregrounds, in particular from the cosmic structure hosting the VHE source (e.g. the Virgo cluster for M87), those permeating our Milky Way, and those produced by the galaxy hosting the nuclear source.
Some brief discussions and the conclusions appear in Sec.\ref{S5}.

A standard cosmology is adopted,
with $H_0 \simeq 70 \, {\rm km} \, {\rm s}^{-1} \, {\rm Mpc}^{-1}$ being the Hubble constant,  ${\Omega}_M \simeq 0.3$ and ${\Omega}_{\Lambda} \simeq 0.7$ are the average cosmic density of matter and dark energy, respectively.


\section{Estimating the gamma-gamma interaction optical depth}
\label{S2}

We detail here how we evaluate the optical depth $\tau_{\gamma\gamma}$ for photons with observed energy $E_0$ produced by an astrophysical source placed at redshift $z_s$ and interacting with background photons (EBL or local photon foregrounds) with energy $\epsilon$ through the pair production process $\gamma\gamma \to e^+e^-$. The emitted photon energy $E$ is linked to the observed one by $E=(1+z_s)E_0$ due to cosmological expansion.

If $\varphi$ is the angle between the directions of incoming photons (from the source and from the background), the probability of the $\gamma\gamma \to e^+e^-$ interaction process is described by the Breit-Wheeler cross-section $\sigma_{\gamma \gamma}(E,\epsilon,\varphi)$~\cite{breitwheeler,heitler} reading
\begin{eqnarray}
\label{crossSecGG}
&\displaystyle\sigma_{\gamma \gamma}(E,\epsilon,\varphi)  = \frac{2\pi\alpha^2}{3m_e^2} \left(1-\beta^2 \right) \times \,\,\,\,\,\,\,\,\,\,\,\,\,\,\,\,\,\,\,\,\,\,\,\,\,\,\,\,\,\,\,\,\,\,\,\,\,\,\,\,\,\,\,\,\,\,\,\,\, \nonumber \\
&\displaystyle \,\,\,\,\,\,\,\,\,\,\,\,\, \times  \,  \left[2 \beta \left( \beta^2 -2 \right) + \left( 3 - \beta^4 \right) \, {\rm ln} \left( \frac{1+\beta}{1-\beta} \right) \right]~,
\end{eqnarray}
with
\begin{equation} 
\label{betaGG}
\beta(E,\epsilon,\varphi) \equiv \left[ 1 - \frac{2 \, m_e^2 \, c^4}{E \epsilon \left(1-\cos \varphi \right)} \right]^{1/2}~, 
\end{equation}
where $\alpha$ is the fine-structure constant, $c$ is the light speed and $m_e$ the electron mass. From Eq.~(\ref{betaGG}) we infer that the process is kinematically allowed for $\beta^2 >0$, which implies that, regarding $E$ as an independent variable, the background photon energy $\epsilon$ must satisfy the inequality
\begin{equation} 
\label{ethr}
\epsilon > {\epsilon}_{\rm thr}(E,\varphi) \equiv \frac{2 \, m_e^2 \, c^4}{ E \left(1-\cos \varphi \right)}~,
\end{equation}
in order for the process to occur.

The cross section $\sigma_{\gamma \gamma}(E,\epsilon,\varphi)$ is maximized at the value ${\sigma}_{\gamma \gamma}^{\rm max} \simeq 1.70 \times 10^{- 25} \, {\rm cm}^2$ for $\beta \simeq 0.70$. Considering head-on collisions ($\varphi = \pi$), $\sigma_{\gamma \gamma}(E,\epsilon,\pi)$ is maximal for background photons with energy 
\begin{equation} 
\label{epsHeadOn}
\epsilon (E) \simeq \left(\frac{500 \, {\rm GeV}}{E} \right) \, {\rm eV}~.
\end{equation}
In the case of an isotropic background photon distribution, the maximal cross section is attained when background photons possess energy~\cite{gouldschreder1967}
\begin{equation} 
\label{EpsIso}
\epsilon (E) \simeq \left(\frac{900 \, {\rm GeV}}{E} \right) \, {\rm eV}~,
\end{equation}
equivalent to Eq. \ref{eq1}.

We are now in a position to evaluate the optical depth $\tau_{\gamma\gamma}$ following the derivation by~\cite{nishikov1962,gouldschreder1967,faziostecker1970}. We treat the case of interaction with local background photons and that with the EBL, separately.

In the case of interaction with local background photons, we can neglect the cosmological evolution. Then, the optical depth $\tau_{\gamma\gamma}$ for source photons interacting with local background photons over a distance $d$ reads
\begin{eqnarray}
\label{tauLoc}
\tau_{\gamma\gamma}(E, d) = \int_0^{d} {\rm d} x \, \int_{-1}^1 {\rm d}({\cos \varphi}) ~ \frac{1- \cos \varphi}{2} \ 
\times \nonumber \\
\times  \, \int_{\epsilon_{\rm thr}(E ,\varphi)}^\infty  {\rm d} \epsilon \, n_{\gamma}(\epsilon, x) \,  
\sigma_{\gamma \gamma} ( E, \epsilon, \varphi )~, \ \ 
\end{eqnarray}
where $n_{\gamma}(\epsilon, x)$ is the local photon background spectral number density, which, in the general case, shows a dependence on the distance in the direction denoted by $x$. Finally, calling $z_s$ the redshift of the source, we can use the relation $E=(1+z_s)E_0$ to express $\tau_{\gamma\gamma}$ in terms of the observed energy $E_0$.

In the case of interaction with the EBL, we must take into account the cosmological expansion. Then, the optical depth $\tau_{\gamma\gamma}$ for photons emitted at redshift $z_s$ interacting with the EBL photons with spectral number density $n_{\gamma}(\epsilon[z], z)$ reads
\begin{eqnarray}
\label{tauExt}
\tau_{\gamma\gamma}(E_0, z_s) = \int_0^{z_s} {\rm d} z ~ \frac{{\rm d} l(z)}{{\rm d} z} \, \int_{-1}^1 {\rm d}({\cos \varphi}) ~ \frac{1- \cos \varphi}{2} \ 
\times \nonumber \\
\times  \, \int_{\epsilon_{\rm thr}(E(z) ,\varphi)}^\infty  {\rm d} \epsilon(z) \, n_{\gamma}(\epsilon(z), z) \,  
\sigma_{\gamma \gamma} ( E(z), \epsilon(z), \varphi )~, \ \ 
\end{eqnarray}
where the distance traveled by a photon in the unit of redshift at redshift $z$ is expressed by
\begin{equation}
\label{lunghez}
\frac{{\rm d} l(z)}{{\rm d} z} = \frac{c}{H_0} \frac{1}{\left(1 + z \right) \left[ {\Omega}_{\Lambda} + {\Omega}_M \left(1 + z \right)^3 \right]^{1/2}}~.
\end{equation}
%

   \begin{table*}[!htbp]
      \caption[]{List of best VHE targets for IR-EBL measurements.}
         \label{table1}
\begin{tabular}{ c c c c c  }        
            \hline
            \noalign{\smallskip}
            \hline
Target & Coordinates & Distance & $F_{10 \, \rm TeV}$ $^{\mathrm{a}}$    & Max energy [TeV]    \\
 name  & RA, DEC     & [Mpc]    &   ($\epsilon^2 dn/d\epsilon )$  &  for CTAO-S          \\						
       &   (2000)    &          &    $[\rm erg/cm^2/sec]$             &  100 h exposure     \\
            \noalign{\smallskip}
            \hline
						\noalign{\smallskip}
   M87     & 12h30m49.4233s, +12d23m28.043s & 16.5    & $1.1\ 10^{-12}$ &  $\sim$60    \\
	 IC 310  & 03h16m42.9775s, +41d19m29.899s & 58.6    & $4\ 10^{-11}$ & $\sim$60    \\
            \noalign{\smallskip}
            \hline
             \noalign{\smallskip}
  Mkn 501     & 21h09m23.0475s, -01d50m16.347s  & 133     & $10^{-10}$ $^{\mathrm{b}}$          &  30    \\
 Centaurus A  & 13h25m27.6150s, -43d01m08.806s  & 3.8    & $6\ 10^{-14}$                         &  20    \\
 NGC 4278     & 12h20m06.8253s, +29d16m50.713s  & 16.4   & $5\ 10^{-13}$ $^{\mathrm{c}}$         &  $-$    \\
 NGC 1275     & 03h19m48.1601s, +41d30m42.103s  & 60     & $\approx 2\ 10^{-11}$ $^{\mathrm{d}}$ &  $-$    \\
            \noalign{\smallskip}
            \hline
         \end{tabular}
\begin{list}{}{}
\item[$^{\mathrm{a}}$] Approximate flux during high states or a flare.
\item[$^{\mathrm{b}}$] Approximate flux during an historical flare in 1996 \cite{2001A&A...366...62A}.
\item[$^{\mathrm{c}}$] Approximate flux measured by the LHAASO observatory in \cite{2024ApJ...971L..45C}.
\item[$^{\mathrm{d}}$] Approximate flux during a set of flaring events in 2017 as reported in \cite{2017ICRC...35..662G}.
\end{list}
   \end{table*}

\section{A test case: constraining the IR Extragalactic Background}
\label{S3}

\subsection{EBL in the optical/near-IR}

Extragalactic photon backgrounds, like the cosmic microwave radiation and the EBL, as well as those well measured in X-rays and gamma-rays, are fundamental observables for cosmology because they constitute the repositories of all radiant energy produced by cosmic sources and structures since the Big Bang. 
The three main physical processes generating this energy (and radiation) include thermonuclear reactions in stars, gravitational accretion in Active Galactic Nuclei (AGN), and possibly also populations of decaying particles, more or less diffuse in space, for example generated during the early phases of cosmic expansion (this still speculative of course). The latter might even constitute the Dark Matter required by astrophysics and cosmology.
Therefore, these background radiations offer essential data to understand how the Universe has taken shape and evolved.

Thanks to more than 60 years of worldwide observational efforts mostly from space, the radiation intensities are currently well measured over a large fraction of the 19 orders of magnitude in photon energy ($10^{-7}$ to $10^{12}$ eV) of the whole e.m. spectrum.
It is an unfortunate, but inevitable, coincidence that such radiations cannot be directly measured where they would be particularly informative, that is in the so-called EBL range, 0.1 to 300 $\mu {\rm m}$. This is indeed where cosmic radiating sources would be mostly expected to emit, but also where enormous foregrounds from air, the solar system, and the Galaxy overwhelm any extragalactic components.

The sources of the EBL are both point-like unresolved objects, like distant galaxies and AGNs, and more diffuse structures and components, like could originate from more or less homogeneous distributions of low-luminosity objects (e.g. primeval Population III stars), or even decaying elementary particles, or low-luminosity undetectable halos around galaxies.
While the contributions to the EBL by resolved sources can be estimated by integrating sufficiently deep number counts, any diffuse components would remain undetectable.
This motivated Aharonian et al. \cite{2006Natur.440.1018A} 
to exploit VHE observations of blazars at moderate redshifts to constrain the EBL in the optical to near-IR via the photon-photon interaction process, obtaining the EBL photon density $n_\gamma$ from Eqs. \ref{eq1a} and \ref{eq2}. 
Their result was that the total EBL there cannot significantly exceed the resolved source emissions, thus ruling out same claimed large excesses by very-high-$z$ emitters, that turned out to be incorrect subtractions of the Zodiacal light. This is confirmed by Postman et al. \cite{2024ApJ...972...95P} based on unique data by the New Horizon spacecraft that allowed them to largely exclude the Zodiacal contamination in the optical by observing from the outer solar system.

\subsection{EBL in the mid- and far-IR}

The situation at the longer IR wavelengths of the EBL, inaccessible by direct detection due to the huge Interplanetary Dust (IPD) and Galactic emissions from 3 to $>100\ \mu {\rm m}$, would take advantage by similar studies based on the gamma-gamma opacity measurements, except that, following Eq. \ref{eq1}, going longward in wavelength requires moving the spectral analysis to the highest VHE energies.
To obtain significant constraints on the IR-EBL up to its expected maximum around $\lambda \simeq 100\ \mu {\rm m}$ (e.g. Franceschini et al. \cite{2008A&A...487..837F,2017A&A...603A..34F}, henceforth FR2017) would require spectral measurements for VHE sources around 50 TeV or more.
As mentioned in Sec. \ref{S1}, this may become possible with the new-generation instrumentation, particularly the ASTRI Mini Array, CTAO-South, and the Large Array of Cherenkov Telescopes (LACT) at the LHAASO site \cite{ShoushanZhang}.

The critical choice of the target sources for this task, however, requires to balance their distance and luminosity against the chance to detect the highest energy VHE photons.
The extreme high-frequency peaked BL-Lac (EHBL) objects (e.g. \cite{2019MNRAS.486.1741F}), potentially the natural targets having the highest peak synchrotron frequency and the hardest spectra in the Fermi database, are too faint and too distant to be detected at high VHE energies \cite{2019AA...629A...2F}.


Sources at much lower distances are then required. The preliminary list of the best potential targets reported in \cite{2019AA...629A...2F,2022JHEAp..35....1V} is updated in Table \ref{table1}. The top section of the table contains our two top priority sources, the lower one some other objects of interest.
As for the latter, MKN 501 and Centaurus A offer an illustration of why the distance is so critical.
MKN 501 experienced in 1996 an historical outburst during which the source flux exceeded that of the Crab, with a very hard spectrum and no sign of convergence up to the highest-energy detected photons of $\sim$20 TeV, an event never repeated later. In spite of this and even assuming the extreme sensitivity of CTAO-South and a very long campaign, the source could not be detected above $\sim$30 TeV due to the EBL by known-sources and its large distance of 133 Mpc.
At the other extreme, Centaurus A is so close that IR-EBL effects could start being observable only at several tens of TeV, however at flux levels that will not be reached by the most sensitive arrays due to its low luminosity.

   \begin{figure}
   \centering
   \includegraphics[width=9cm]{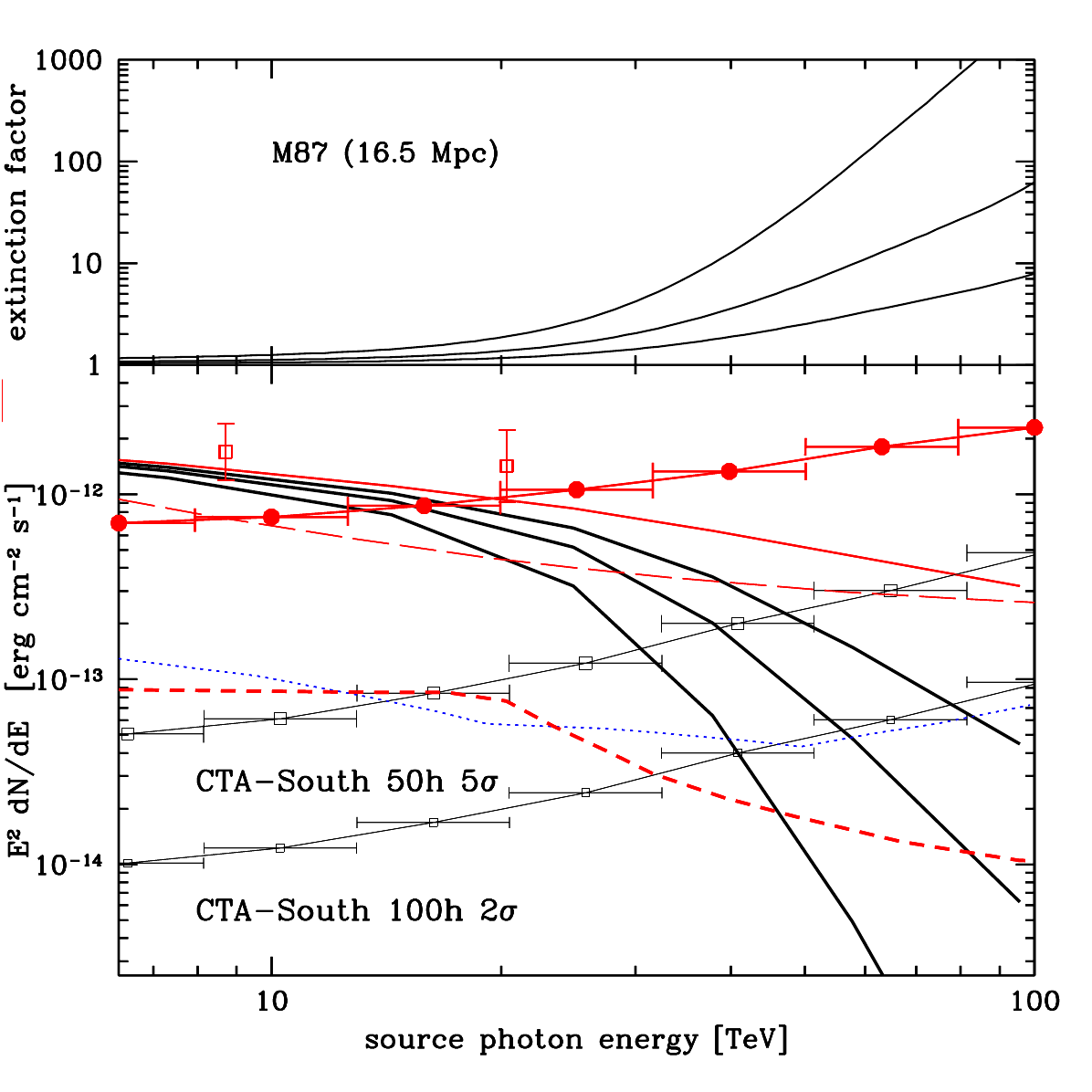}
\caption{
\textit{Top panel: }  The intermediate line is the photon-photon absorption correction ($\exp[\tau_{\gamma\gamma}]$)
for the source M87 at 16.5 Mpc distance, based on the EBL model by FR2017. The upper and lower lines correspond to a factor two higher and lower EBL intensity, respectively.
\textit{Bottom panel:} The observed (open red continuous line) and absorption-corrected (black lines) spectral data are reported corresponding to the three EBL intensities in the above panel. 
The 50-hour $5\sigma$ and 100 hours $2\sigma$ sensitivity limits for CTAO-South, and the HAWC 5 years limits are shown (long red dashed). The blue dotted line and the red short dashed one indicate the SWGO and LHAASO 5 years $5\sigma$ limits, respectively. The red upper curve is the sensitivity limit of the ASTRI Mini Array 50h 5$\sigma$ \cite{2022JHEAp..35....1V}. This updates a similar figure in \cite{2019AA...629A...2F} based on a wrong distance measurement.
}               
\label{M87} 
   \end{figure}

   \begin{figure}
   \centering
   \includegraphics[width=9cm]{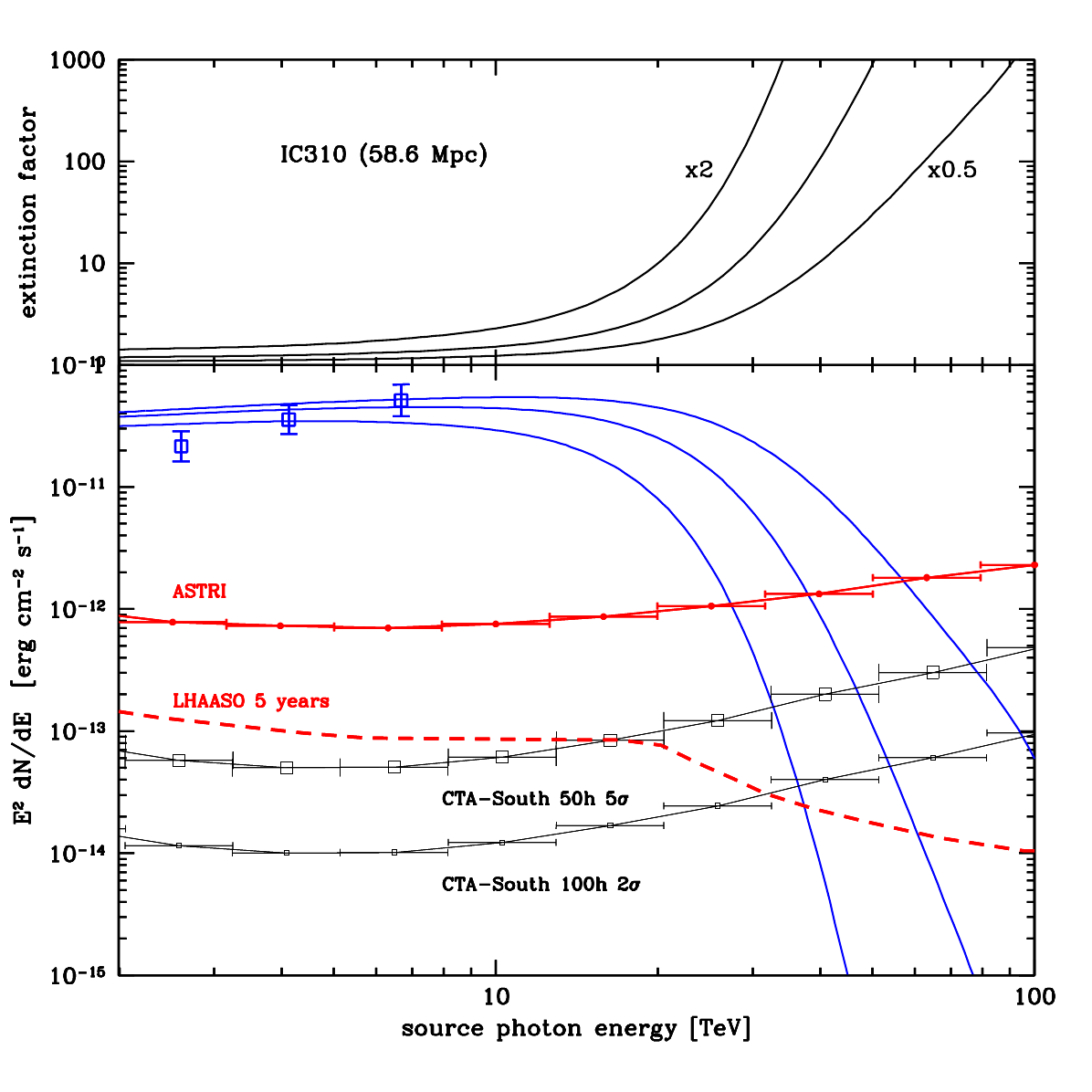}
\caption{
\textit{Top panel:} The intermediate line is the photon-photon absorption correction ($\exp[\tau_{\gamma\gamma}]$) for the source IC 310 at the 58.6 Mpc distance, based on the EBL model by FR2017. The upper and lower lines correspond to a factor two higher and lower EBL intensity, respectively.
\textit{Bottom panel:} The blue data-points were taken during an outburst phase, the continuous lines are corrected for the three EBL models as in the panel above.
The 50-hour $5\sigma$ and 100 hours $2\sigma$ sensitivity limits for CTAO-South is shown. The red dashed line indicates the LHAASO 5 years $5\sigma$ limits, while the red upper curve is the sensitivity limit of the ASTRI Mini Array 50h 5$\sigma$ \cite{2022JHEAp..35....1V}.
}               
\label{IC310} 
   \end{figure}

The AGN in NGC 4278 has been recently detected by LHAASO at appreciable TeV flux \cite{2024ApJ...971L..45C} and is located at the same suitable distance of M87 (although not sharing the same celestial position, see table), while NGC 1275 (3C 84), recently detected by various VHE observatories, is in the same Perseus cluster as IC310 and can be observed by IACT arrays during the same long integrations.

Coming to our two optimal targets, M87 and IC310, they display the best combination of luminosity and distance to allow detection of the highest energy VHE photons on the tens of Mpc extragalactic scale. Note that with respect to the analysis in \cite{2019AA...629A...2F} we had to significantly update the source distances in accordance to the most recent measurements (NASA/IPAC Extragalactic Database \cite{footnote1}), this having a significant impact on the photon-photon opacity corrections.
In addition to the mentioned tests for new physics and the standard model of fundamental interactions, spectral analyses at tens of TeV are needed to constrain the EBL at the longest IR wavelengths.
To this end, we report in Figs. \ref{M87} and \ref{IC310} a simulation of the effects of factor-two variations of the IR-EBL intensity with respect to the model by FR2017. A lower value of the IR-EBL has been recently indicated in \cite{2023SciA....9J2778C} to try to explain a detection of photons up to 13 TeV from a GRB at $z=0.151$, and we modeled this with the bottom curve in the top panels of the figures.
Assuming instead that the IR-EBL from resolved-sources is as in FR2017 and that a diffuse component, like due to diffuse dust or decaying particles, adds to that, we modeled the $\gamma\gamma$ absorption from such a background as the top line in the top panels and the lower EBL-absorbed curves in the bottom (see \cite{2019AA...629A...2F} for more information on the model curves for the two sources).

The message from these simulations is that important constraints on the IR-EBL can be obtained from sufficiently long VHE measurements during flaring or high states of sources in Table \ref{table1}, assuming that we can rely on valid models of the intrinsic source emission.

   \begin{figure}
   \centering
   \includegraphics[width=9cm]{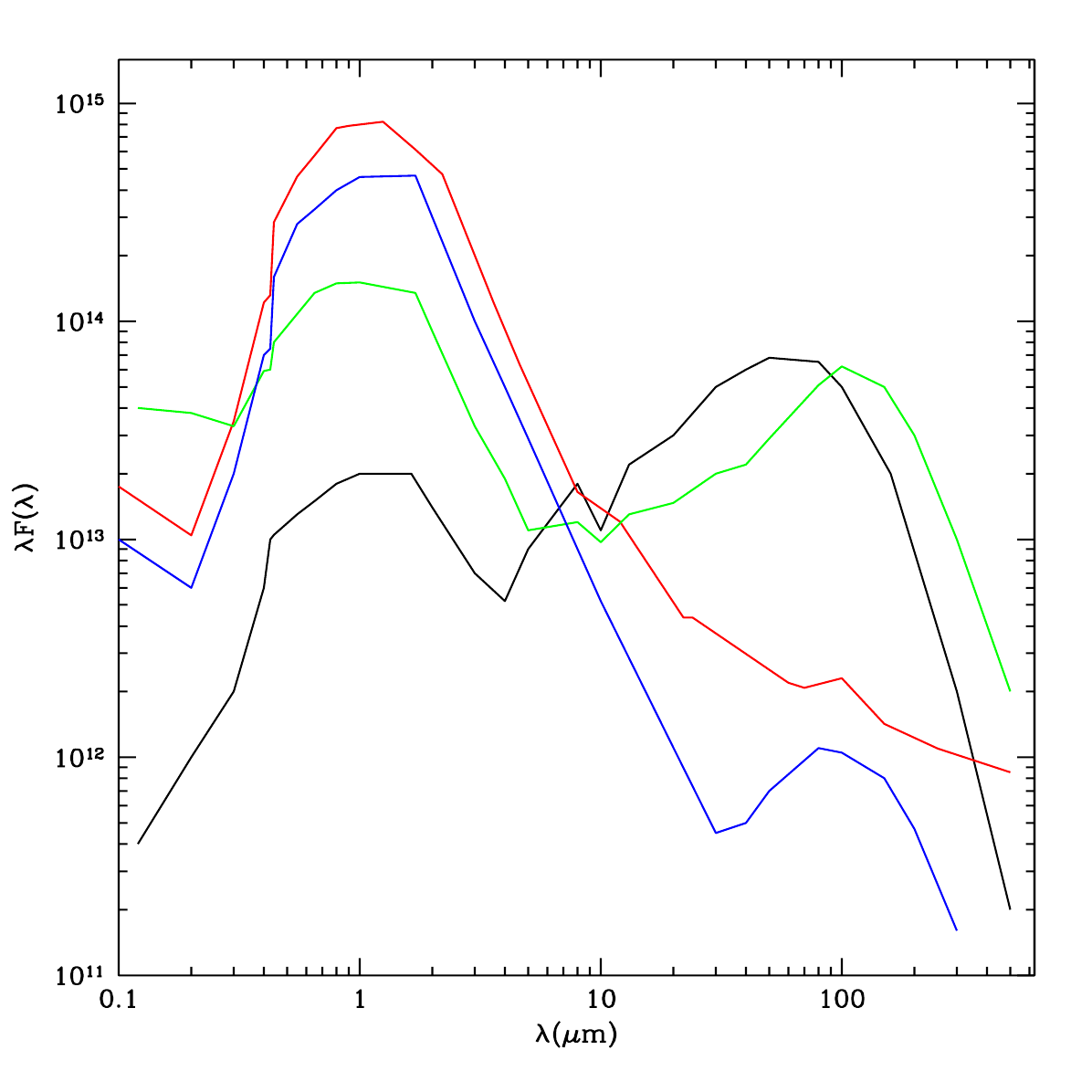}
\caption{
Renormalized spectral energy distributions for galaxy populations in our spectral analysis. 
The $y$-axis is in $\lambda F(\lambda)=\nu F(\nu)$ units [e.g. erg cm$^{-2}$ sec$^{-1}$].
The black line there is the average SED of a luminous infrared galaxy, The blue line corresponds to the model SED of an elliptical galaxy, the green line fits the SED data of the Sbc spiral galaxy NGC 6946. 
The red SED fits the broad-band data of M87. See text for more details.
}               
\label{spectra} 
   \end{figure}

\section{The optical depths by the local radiation foregrounds}
\label{S4}

Because we are dealing with VHE sources at moderate to low distances, the gamma-gamma interaction and pair production effects have to deal, in addition to the EBL, with local foregrounds that the VHE photons have to cross before detection.
The various contributions to these foregrounds are discussed in the following sections.

\subsection{Radiations from the host cluster: the cases of the Virgo (M87) and Perseus (IC310, NGC1275) clusters}
\label{S41}

One of our favorite sources, M87, is resident close to the center of the Virgo cluster, the most conspicuous cosmic structure in our vicinity. We have modeled the diffuse light in the cluster by integrating the emissions from galaxies reported in the Extended Virgo Cluster Catalogue (EVCC) \cite{2014ApJS..215...22K}. This is a large complete database selected from the Sloan Digital Sky Survey (SDSS, Data Release 7) that covers an area of 60 Mpc$^2$ at the Virgo distance, i.e. including the entire cluster in excess. We have selected all galaxies within a radius $R$ of 3 degrees of M87's position, assumed to be the cluster center,  within a radial velocity of 1000 $\rm km/sec$ (the cluster line-of-sight velocity dispersion), and differentiated them for the morphological type between early- (ellipticals/S0) and late-types (spiral/irregular). 

The integrated surface brightness $I_\nu$ and the cluster photon density $n_\gamma$ are then calculated in the EVCC $r$ band that we established as our spectral reference:
\begin{equation}\label{eq4}
 I_{\nu}=  \sum{F_{r,i}} / \pi R^2		, 
\end{equation}
with $F_{r,i}=R_{\rm calib}\times 10^{-0.4 \, r_i}$ the flux of the $i-th$ EVCC galaxy, $R_{\rm calib}=3080\ \rm Jy$ for the $r$ band, and
\begin{equation}\label{eq5}
 n_\gamma=4\pi I_{\nu}/c .
\end{equation}
For simplicity, in our calculations of the radiation fields of cluster origin we assume an homogeneous source spatial distribution within the cluster radius.

These $r$-band $n_\gamma$ photon densities are then extrapolated to all other wavebands from UV to the millimeter using synthetic spectral energy distributions (SED) of various source populations, that are plotted in Fig. \ref{spectra}.
The black line there is the average SED of a luminous infrared galaxy, that we have modeled from the study \cite{1998ApJ...509..103S} of the starburst galaxy M82.
The blue line corresponds to the model SED of an elliptical galaxy by the same authors, as well as the green line fitting the SED data of the Sbc spiral galaxy NGC 6946 as representative of the late-type galaxy spectra. 
Finally, the red SED fits the broad-band data of M87, as discussed later in Sec. \ref{S42}.
Having split the galaxy contributions to the light in Virgo into the early-type and late-type components, we have extrapolated the corresponding $r$ band photon density fields with the elliptical and spiral SED in the figure, respectively.

   \begin{figure*}{}
   \centering
   \includegraphics[height=11cm,width=15cm]{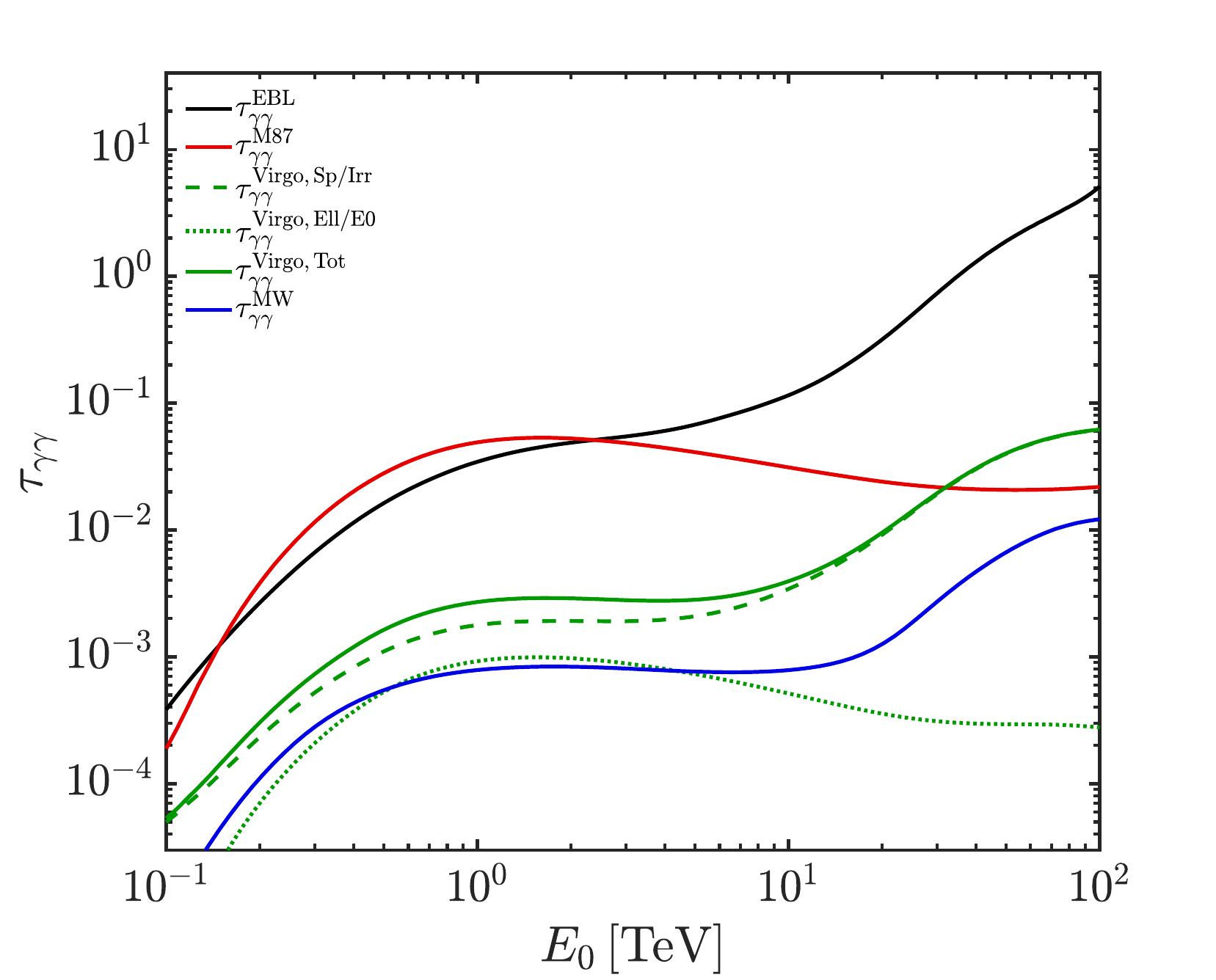}
\caption{
Optical depths $\tau_{\gamma\gamma}$ as a function of photon energy for the VHE emission from M87 due to $\gamma\gamma$ interactions with various radiation components.
The black line is based on the EBL model by FR2017 at the M87 distance of 16.5 Mpc.
The continuous green line is due to the diffuse light in Virgo (dotted and dashed ones are for the early- and late-type galaxy contributions, see Sec. \ref{S41}). The red curve is the optical depth due to the radiation field internal to the source (Sec. \ref{S42}). The blue line is the absorption due to photons in the Milky Way (Sec. \ref{S43}).
}               
\label{opacityM87} 
   \end{figure*}

A further important factor to consider was how much entirely diffuse light may be originated by diffuse components inside the cluster, like may be stellar populations lost by the cluster galaxies or diffuse dust and free electrons scattering the local background photon field. This diffuse intra-cluster light (ICL) in galaxy clusters has been the subject of various analyses (e.g. \cite{Zibetti2005,Gonzalez2007,Burke2015,MontesTrujillo2018,2019ApJ...874..165Z} to mention a few), with contributions of the ICL found to range from $\sim10\%$ to up to 50\% of the total cluster light.
The tendency of all these studies is also to find a marked dependence of the ICL content on the cluster richness and age, presumably due to the effects of dynamical friction spoiling part of the stellar content of the cluster galaxies as the cluster evolves. For a young, irregular, and diffuse cluster like Virgo we then assumed a low value for ICL of $12\%$ of the total cluster light, independent of wavelength.

Once calculated the foreground photon fields, we used them to compute the photon-photon opacities for VHE gamma-rays emitted by M87, according the recipes in Sec.  \ref{S2}. The results appear in Fig. \ref{opacityM87} as the green dotted line for the Virgo early-type component and dashed line for the late-types, with their sum shown as the continuous green line.
These contributions to the foreground $\gamma\gamma$ opacities are compared to the opacity from FR2017 due to the EBL reported as the continuous black line, that keeps far above the former at all VHE energies.

   \begin{figure*}{}
   \centering
   \includegraphics[height=11cm,width=15cm]{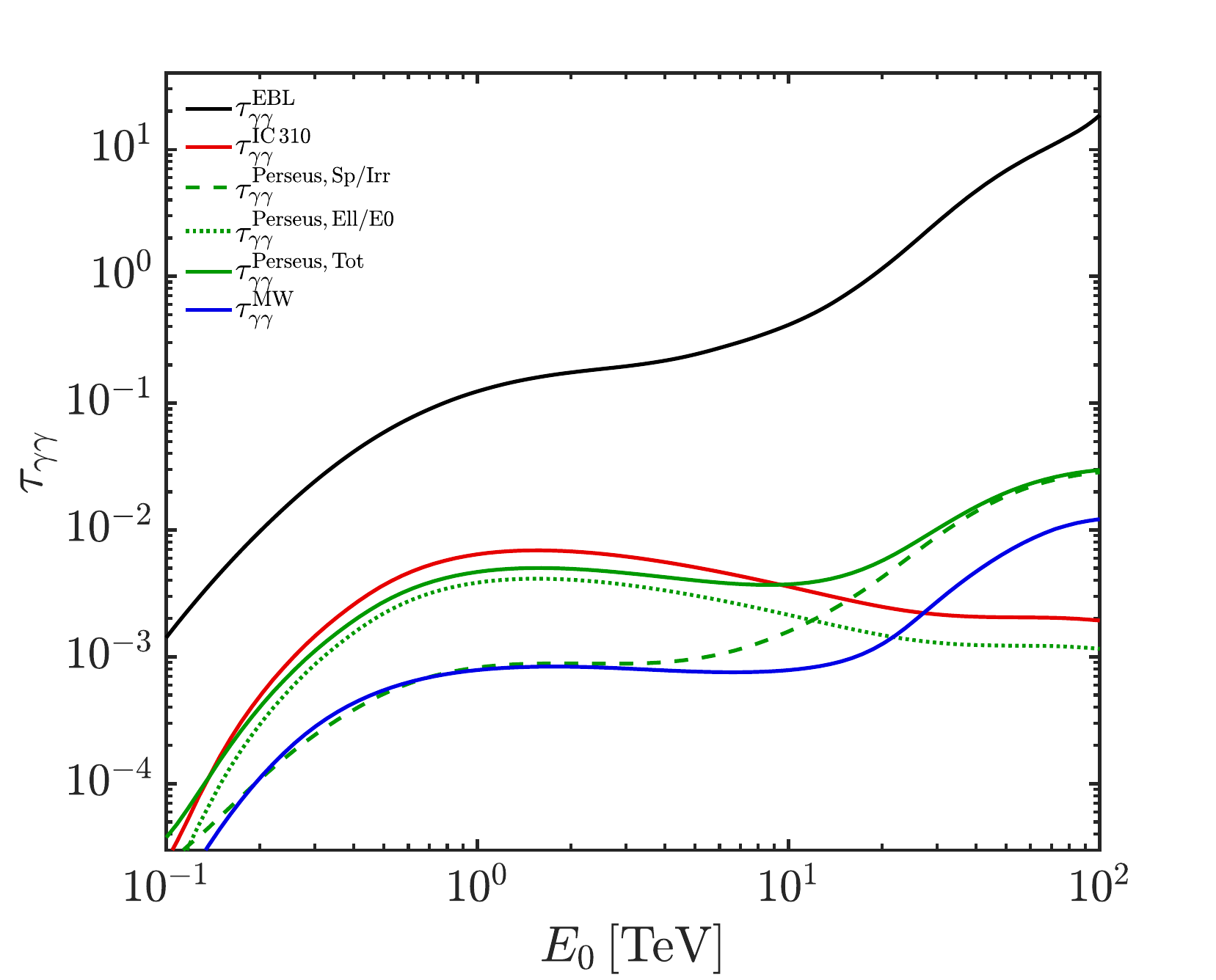}
\caption{
Optical depths $\tau_{\gamma\gamma}$ as a function of photon energy for the VHE emission from IC310 due to $\gamma\gamma$ interactions with various radiation components.
The black line is based on the EBL model by FR2017 at the IC310 distance of 58.6 Mpc.
The continuous green line is due to the diffuse light in Perseus (dotted and dashed ones are for the early- and late-type galaxy contributions, see Sec. \ref{S41}). The red curve is the optical depth due to the radiation field internal to the source (Sec. \ref{S42}). The blue line is the absorption due to photons in the Milky Way (Sec. \ref{S43}).}               
\label{opacityIC310} 
   \end{figure*}

Perseus, hosting both IC310 and NGC 1275, is a very rich, massive and regular low-redshift cluster. In such a case, to estimate the contribution by the cluster radiation field to the $\gamma\gamma$ opacities we have simplified our analysis with the following procedure. We have considered an average photon density in the cosmic field in V-band of $10^8\ \rm L_\odot/Mpc^3$ corresponding to the average galactic cosmic number density of 1 galaxy$/30\ \rm Mpc^{-3}$ and scaled this photon field density to that inside the cluster by a factor $\sim 900$. This comes from assuming a galaxy density of $\sim 30$ galaxies per cubic Mpc inside the cluster -- that is $\sim 2000$ galaxies inside a typical rich-cluster sphere of 2.5 Mpc radius -- times 30 cubic Mpc per galaxy in the field. As a conservative assumption about the galaxy content and spectral corrections from V-band to the rest of the spectrum, we have assumed that 20\% \ of the galaxies are spirals/irregulars and adopted for them the SED of a spiral galaxy as in Fig.\ref{spectra}, 
while the remaining 80\% \ are ellipticals with their own spectral properties.

To evaluate the ICL in this rich cluster of galaxies we have referred to \cite{2019ApJ...874..165Z} that performed a very extensive analysis of $\sim 300$ galaxy clusters in the redshift range of $0.2-0.3$ based on the Dark Energy Survey (DES). 
They estimate that ICL contributes $44\% \pm 17\%$ of the total cluster luminosity within 1 Mpc. To account for ICL in Perseus, we then multiplied the galaxy contribution by a constant factor 1.78.

Again the procedure in Sec. \ref{S2} was used to calculate the cluster contribution to the photon-photon opacities, and compared them to the EBL optical depths for IC310 in Fig. \ref{opacityIC310}. There again the EBL is the black line and the estimated contributions of the cluster are the green ones.

\subsection{Radiations from the host galaxy: the cases of M87, IC310, and Centaurus A}
\label{S42}

A second foreground radiation component to consider is generated by stars and the interstellar medium in the galaxy hosting the VHE emitter.
In the sample that we considered in Table \ref{table1}, the latter tend to be massive galaxies of the early morphological type hosting active super-massive black-holes and jets directed close to our line-of-sight.
The galaxy hosting the VHE emitter in M87 is a particularly important case, because it is an ultra-massive very extended cD galaxy in the core of Virgo, that deserves a special attention.
To this end we have used the deep surface photometry in the R Johnson band by \cite{2005AJ....129.2628L} from the galaxy core out to 2000 arcsec, corresponding to 160 kpc at the galaxy distance: this is one of the most extended galactic stellar halos known.
We have applied spectral corrections to this surface brightness distribution using a SED that from 0.1 to 0.8 $\mu {\rm m}$ is based on the model spectrum of an elliptical galaxy from \cite{1998ApJ...509..103S}. From 0.8 to 500 $\mu {\rm m}$ we have used the broad-band integrated photometry from the NASA/IPAC Extragalactic Database \cite{footnote1}, as it appears to show an excess flux at $\lambda >10\ \mu {\rm m}$ with respect to that of a standard quiescent elliptical.
The $\gamma\gamma$ opacity due to this radiation component, calculated with the recipes in Sec. \ref{S2}, is reported as the red continuous line in Fig. \ref{opacityM87}.

IC310 shows the morphology of a standard early-type S0 galaxy with a total size in the $K_s$ band of 104 arcsec from 2MASS and a total visual magnitude of $\rm V=8.164$ (from the US Naval Observatory, see \cite{2015ApJ...810...14A}) and $K_s=9.145$ \cite{2003yCat.7233....0S}.
For this host galaxy we have adopted the spectrum of a pure elliptical, as the blue curve in Fig. \ref{spectra}.
The results about the internal opacity in IC310 appear as the red line in Fig. \ref{opacityIC310}.

\begin{figure}
   \centering
   \includegraphics[width=9.05cm]{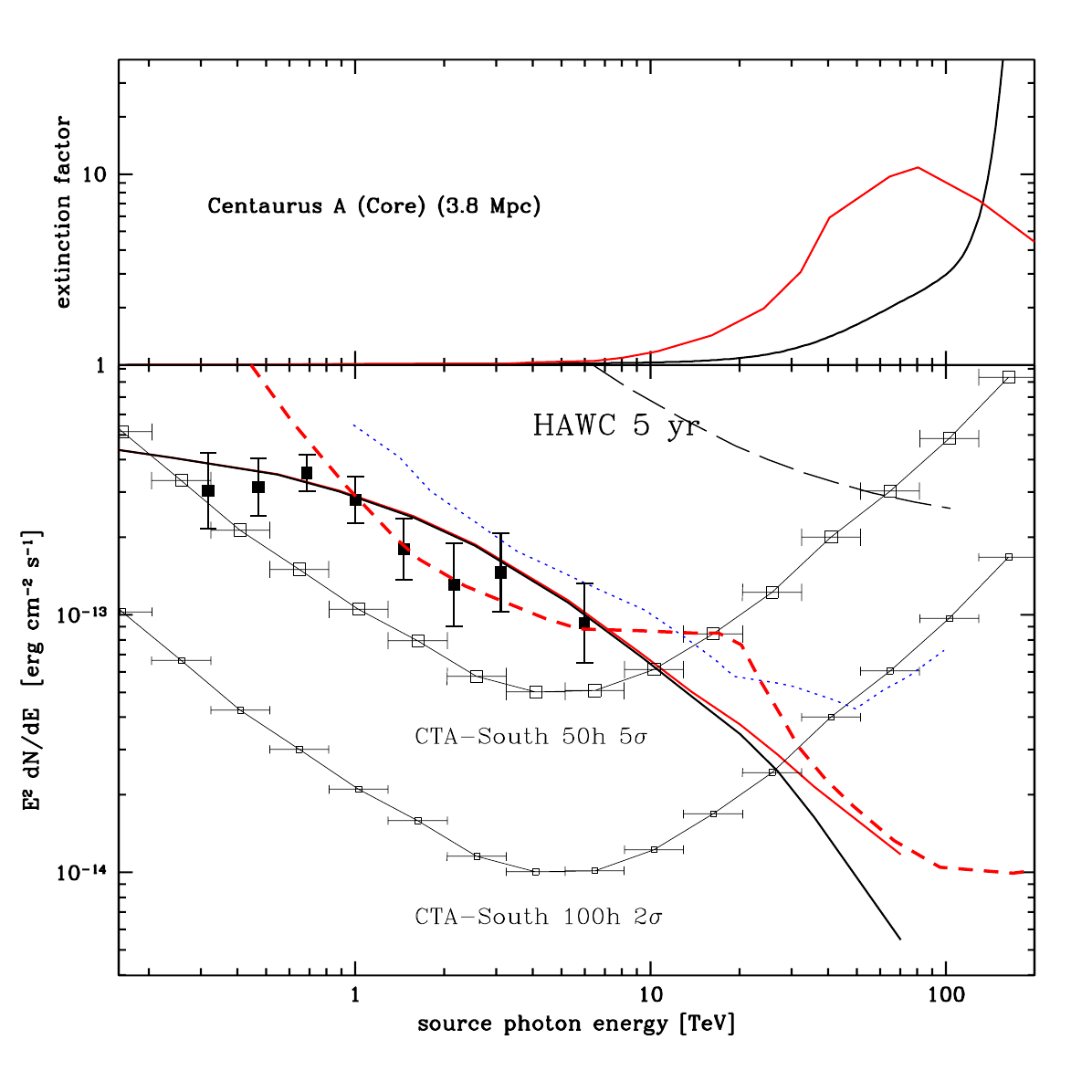}
\caption{
\textit{Top panel:} The black line is the photon-photon absorption correction ($\exp[\tau_{\gamma\gamma}]$) from EBL for the source Centaurus A at the 3.8 Mpc distance, based on the FR2017 model. The red line quantifies the further extinction due to the radiation background intrinsic to the host galaxy.
\textit{Bottom panel:} The red continuous line is the source intrinsic spectrum, and the black line is corrected for EBL absorption only.
The additional absorption intrinsic to the source would essentially cut down the spectrum above $\sim$10 TeV (Sec. \ref{S5}).
See caption to Fig. \ref{M87} for the other data and curves in the figure.
}               
\label{CenA} 
   \end{figure}

Finally, a VHE source of potential interest reported in Table \ref{table1} is the radiogalaxy Centaurus A, the AGN closest to us. This is a peculiar object consisting in a massive elliptical deeply interacting with a spiral galaxy, showing manifest evidence of dusty star formation and dust absorption.
The galaxy and its large stellar halo have been studied in \cite{2022A&A...657A..41R} based on Hubble Space Telescope imaging obtained with the WFPC2, ACS, and WFC3 cameras. 
We have used this surface brightness distribution in the R band to infer the halo photon number density and corrected it spectrally using the SED of a LIRG galaxy like M82 (black line in Fig. \ref{spectra}).
The results for the $\gamma\gamma$ extinction factor internal to the host galaxy are shown as the red line in the top panel of Fig. \ref{CenA}, and compared there to that due to the EBL.

\subsection{Radiations in the Milky Way at high galactic latitudes}
\label{S43}

A further radiative foreground to be considered when observing distant VHE sources is produced by the stellar and dust layers in the Milky Way (MW). This obviously makes a continuum photon distribution with scale-height that is not easy to quantify, as it depends on the direction in the sky.
In the present study we limited ourselves to consider the integrated surface brightness distribution of galactic photons in the direction of the galactic poles at all wavelengths of interest from UV to the millimeter, as reported in the review paper \cite{1998A&AS..127....1L}.
To the integrated faint stellar contribution in the optical, the bright star's one has to be added, which amounts to approximately 20\% of that from faint stars (see their Fig. 61).
An additional contribution that we considered is due to Diffuse Galactic Light (DGL) produced by scattering of stellar photons by dust grains in the interstellar space: this was estimated to make another $\sim20\%$\ of the total integrated light from the Milky Way (see Sec. 11 of \cite{1998A&AS..127....1L}).

The Milky Way contributions to the total $\tau_{\gamma\gamma}$ opacities for our best-guess sources are reported as blue continuous lines in Figs. \ref{opacityM87} and \ref{opacityIC310}.  We see that such contributions to $\tau_{\gamma\gamma}$ in the direction of the galactic poles keep always very low, not typically exceeding the 1\%\ value at any wavelengths. Assuming the approximate dependence of the stellar and dust column density, hence of the optical depth, on the galactic latitude $b$ as $\tau_{\gamma\gamma} \propto \csc (b)$, these values of $\tau_{\gamma\gamma}$ from MW would increase only by 40\%\ on average for sky directions at $b=45^\circ$, and a factor $\sim 2$ at $b=30^\circ$, so remaining negligible in any case.

\section{Discussion and conclusions}
\label{S5}

It is of interest for various physical, astrophysical, and cosmological tests to consider the photon-photon interactions on an extragalactic scale of tens of Mpc, involving VHE sources detectable with the new generation of Cherenkov arrays (CTAO, LACT, SWGO, ASTRI Mini Array, among others).

A specific application that we considered concerns the interaction of VHE photons from \textit{jetted} AGNs with the EBL photons at infrared wavelengths. The latter has remained so far largely unmeasurable, as well as unconstrained via $\gamma\gamma$ opacity effects because it would require the detection of photons above the energy of 10 TeV.
To tackle the IR-EBL, we have reviewed a set of potential VHE targets for long observational campaigns or during source flaring events that we partly analysed in our previous paper \cite{2019AA...629A...2F}. The detection by LHAASO of 5 AGNs \cite{2024ApJS..271...25C} is very promising and demonstrates a realistic possibility of measuring the IR-EBL.

When considering the effects of photon-photon interactions on low-redshift sources, however, one important factor that we need considering in our analysis concerns the contributions of the opacities due to local radiation fields -- local \textit{foregrounds} -- that the VHE photons have the chance to cross before reaching our telescopes.
We identified three of such sources of radiation: radiations from the cosmic structure (typically a cluster) hosting the VHE emitter, from the galaxy itself hosting the emitter, and from our own system of reference (the Milky Way).
The results are shown as different colored lines in Figs. \ref{opacityM87}, \ref{opacityIC310}, and \ref{CenA}.

\smallskip\textit{The Milky Way.  \ } To start with, the Milky Way, when looking outside the Galactic disk, $\left|b\right|\gtrsim 20^\circ$, is never expected to cause opacities greater than $1\%$ of the total opacity, that is completely negligible: this consideration also applies to Centaurus A, which lies just below the soft limit of $20^\circ$. For observations of galactic sources, instead, the situation may be different, and quite more complex to evaluate of course. However, the fact that several PeVatron sources have been detected by LHAASO \cite{2024ApJS..271...25C} indicates that such opacities may not be huge at least above 100 TeV, also considering that the Galactic radiation field is expected to fade away at $\lambda>200 \, \mu {\rm m}$.

\smallskip\textit{The host clusters.  \ } Similarly, the two galaxy clusters hosting our two best-guess sources, Virgo for M87 and Perseus for IC310, in neither case do appear to generate such levels of foreground radiation to compromise observations at the highest VHE energies. The difference in the results for the two clusters around 100 TeV are due to differences in the structure and galactic content of the two: we assumed that the Virgo young irregular cluster is dominated by late-type galaxies and attributed them a spiral-galaxy spectrum. For the rich and regular Perseus cluster the only 20\% \ of late-type galaxies makes the cluster contribution to keep lower than 1\% \ of the total opacity at all energies.
This in any case does not change our above conclusion.

\smallskip\textit{The host galaxy of Centaurus A: NGC 5128.  \ } 
One of the VHE sources that we deemed as interesting for the IR-EBL analysis, Centaurus A, is hosted by a massive, luminous and extended galaxy (NGC 5128) responsible for a substantial photon foreground and source of opacity, as illustrated in Fig. \ref{CenA}. Because of its proximity, the optical depth due to the EBL is negligible up to a few tens of TeV, while above $\sim 10$ TeV the $\gamma\gamma$ opacity is dominated by the IR dust emission by the host galaxy.
In conclusion, we expect that the rapid increase of the latter makes this source virtually undetectable above $20-30 \, \rm TeV$.

\smallskip\textit{The host galaxies of M87 and IC310.  \ }
Finally, the two galaxies hosting M87 and IC310 are both massive ellipticals, with stellar masses in the range $M \simeq 10^{11.5 - 12} \, \rm M_{\odot}$. What is peculiar of the host of M87 is its enormous stellar halo extending up to 200 kpc and more,  presumably due to the coalescence of a number of satellite galaxies and stripped stars in the core of the cluster. 
The integrals of this radiation fields in Eq. \ref{tauLoc} then imply a large opacity, quite comparable to that by EBL around 1 TeV but fading down and becoming negligible at higher energies (Fig. \ref{opacityM87}).
On the contrary, for the galaxy hosting IC310, still being a massive elliptical, the lack of an extended halo makes its contribution to always keep very low.

\smallskip

In conclusion, about our planned attempt to constrain the infrared extragalactic background IR-EBL by looking at VHE spectra above 10 TeV for AGNs at few tens of Mpc distances (see \cite{2019AA...629A...2F,2022JHEAp..35....1V}), from our analysis in no case the intervening foreground radiations appeared to compromise the prospects for this attempt.

Concerning our two best-guess targets in Table \ref{table1} and Figs. \ref{M87} and \ref{IC310}, it is interesting to note the difference in the simulated spectra of M87 and IC310: the smoother convergence in the former at the higher energies, the sharper one of the latter, due to the different distances. The consequence is that the inference about the gamma-gamma optical depth is safer for IC310, for which it essentially requires the identification of the well developed exponential cutoff, whereas it is somewhat model-dependent in the former where the EBL cutoff is more difficult to disentangle from the intrinsic spectral behavior of the source.
Other VHE targets at the appropriate distances of several tens of Mpc may show up during the forthcoming campaigns with Cherenkov-light observatories.

\section*{Acknowledgments}
We are grateful to J. Biteau, F. Tavecchio, S. Vercellone, and G. Pareschi for useful comments comments and suggestions.


\begin{thebibliography}{}

\bibitem{addazi2022} A. Addazi {\it et al.}, Prog. Part. Nucl. Phys. {\bf 125}, 103948 (2022).

\bibitem{grRev} G. Galanti and M. Roncadelli, Universe {\bf 8}, 253 (2022).

\bibitem{2006Natur.440.1018A} F. Aharonian {\it et al.}, Nature {\bf 440}, 1018 (2006).

\bibitem{2023SciA....9J2778C} Z. Cao {\it et al.} [LHAASO Collaboration], Science Adv. {\bf 9}, adj2778 (2023).

\bibitem{2022JHEAp..35...52S} S. Scuderi {\it et al.}, J. High Energy Astrophys. {\bf 35}, 52 (2022).

\bibitem{2011ExA....32..193A} M. Actis {\it et al.}, Experimental Astronomy {\bf 32}, 193 (2011).

\bibitem{2019ChA&A..43..457C} Z. Cao {\it et al.}, Chinese Astronomy and Astrophysics {\bf 43}, 457 (2019). 

\bibitem{breitwheeler} G. Breit and J. A. Wheeler, Phys. Rev. {\bf 46}, 1087 (1934).

\bibitem{heitler} W. Heitler, The Quantum Theory of Radiation. Oxford Univ. Press, Oxford (1960).

\bibitem{gouldschreder1967} R. J. Gould and G. P. Schr\'eder, Phys. Rev. {\bf 155}, 1408 (1967).

\bibitem{nishikov1962} A. Nikishov, Sov. Phys. JETP {\bf 14}, 393 (1962).

\bibitem{faziostecker1970} G. G. Fazio and F. W. Stecker, Nature {\bf 226}, 135 (1970).

\bibitem{2024ApJ...972...95P} M. Postman {\it et al.}, Astrophys. J. {\bf 972}, 95 (2024).

\bibitem{2008A&A...487..837F} A. Franceschini, G. Rodighiero and M. Vaccari, Astron. Astrophys. {\bf 487}, 837 (2008).

\bibitem{2017A&A...603A..34F} A. Franceschini and G. Rodighiero, Astron. Astrophys. {\bf 603}, A34 (2017).

\bibitem{ShoushanZhang} S. Zhang, The 8th Heidelberg International Symposium on High-Energy Gamma-Ray Astronomy 2024/09/02-06, Milan, Italy (2024).

\bibitem{2019MNRAS.486.1741F} L. Foffano, E. Prandini, A. Franceschini and S. Paiano, Mon. Not. R. Astron. Soc. {\bf 486}, 1741 (2019).

\bibitem{2019AA...629A...2F} A. Franceschini, L. Foffano, E. Prandini and F. Tavecchio, Astron. Astrophys. {\bf 629}, A2 (2019).

\bibitem{2022JHEAp..35....1V} S. Vercellone {\it et al.}, J. High Energy Astrophys. {\bf 35}, 1 (2022).

\bibitem{2001A&A...366...62A} F. A. Aharonian {\it et al.}, Astron. Astrophys. {\bf 366}, 62 (2001).

\bibitem{2024ApJ...971L..45C} Z. Cao {\it et al.} [LHAASO Collaboration], Astrophys. J. {\bf 971}, L45 (2024).

\bibitem{2017ICRC...35..662G} D. Glawion {\it et al.}, 35th International Cosmic Ray Conference (ICRC2017) {\bf 301} (2017).

\bibitem{footnote1} The NASA/IPAC Extragalactic Database (NED) is funded by the National Aeronautics and Space Administration and operated by the California Institute of Technology.

\bibitem{2014ApJS..215...22K} S. Kim {\it et al.}, Astrophys. J. Suppl. Ser. {\bf 215}, 22 (2014).

\bibitem{1998ApJ...509..103S} L. Silva, G. L. Granato, A. Bressan and L. Danese, Astrophys. J. {\bf 509}, 103 (1998).

\bibitem{Zibetti2005} S. Zibetti, S. D. M. White, D. P. Schneider and J. Brinkmann, Mon. Not. R. Astron. Soc. {\bf 358}, 949 (2005).

\bibitem{Gonzalez2007} A. H. Gonzalez, D. Zaritsky, and A. I. Zabludoff, Astrophys. J. {\bf 666}, 147 (2007).

\bibitem{Burke2015} C. Burke, M. Hilton and C. Collins, Mon. Not. R. Astron. Soc. {\bf 449}, 2353 (2015).

\bibitem{MontesTrujillo2018} M. Montes and I. Trujillo, Mon. Not. R. Astron. Soc. {\bf 474}, 917 (2018).

\bibitem{2019ApJ...874..165Z} Y. Zhang {\it et al.}, Astrophys. J. {\bf 874}, 165 (2019).

\bibitem{2005AJ....129.2628L} Y. Liu {\it et al.}, Astron. J. {\bf 129}, 2628 (2005).

\bibitem{2015ApJ...810...14A} M. Ackermann {\it et al.} [Fermi Collaboration], Astrophys. J. {\bf 810}, 14 (2015).

\bibitem{2003yCat.7233....0S} M. F. Skrutskie {\it et al.}, VizieR Online Data Catalog: 2MASS All-Sky Extended Source Catalog (XSC) (IPAC/UMass, 2003-2006). VizieR Online Data Catalog 7233. VII/233 (2003).

\bibitem{2022A&A...657A..41R} M. Rejkuba, W. E. Harris, L. Greggio, D. Crnojevi{\'c} and G. L. H. Harris, Astron. Astrophys. {\bf 657}, A41 (2022).

\bibitem{1998A&AS..127....1L} Ch. Leinert {\it et al.}, Astron. Astrophys. Suppl. Ser. {\bf 127}, 1 (1998).

\bibitem{2024ApJS..271...25C} Z. Cao {\it et al.} [LHAASO Collaboration], Astrophys. J. Suppl. Ser. {\bf 271}, 25 (2024).


\end{thebibliography}
\end{document}